# Wind-wave stabilization by a foam layer between the atmosphere and the ocean

Yuri M. Shtemler,[1] Ephim Golbraikh,[2] Michael Mond[1]

[1]Department of Mechanical Engineering,

[2]Physics Department,

Ben-Gurion University of the Negev,

P.O.Box 653, Beer-Sheva 84105, Israel

**Abstract**

The study is motivated by recent findings of the decrease in the momentum transfer from strong winds to sea. The Kelvin-Helmholtz instability (KHI) of a three-fluid system of air, foam and water is examined within the range of intermediately short surface waves. The foam layer thickness necessary for effective separation of the atmosphere and the ocean is estimated. Due to high density contrasts in the three-fluid system, even a relatively thin foam layer between the atmosphere and the ocean can provide a significant stabilization of the water surface by the wavelength shift of the instability towards smaller scales. It is conjectured that such stabilization qualitatively explains the observed reduction of roughness and drag.

PACS numbers: 92.60.Cc, 92.10.Fj

**1. Introduction**

Ocean – atmosphere interaction in strong winds is a vital issue, which has been raising recently a growing interest. A linear increase in the momentum transfer from wind to sea waves measured at weak winds is conventionally extrapolated to strong winds (e.g., Large and Pond, 1981). The present study is motivated by recent findings of saturation and even decrease in the drag coefficient (capping) in strong winds starting from $\sim 30\ m/s$, which is accompanied by the production of a foam layer on the ocean surface (see both field and laboratory experiments in Powell et al., 2003 and Donelan et al., 2004). As described in (Donelan et al., 2004) "in very strong winds the character of the ocean surface does change appreciably having intense wave breaking, spume blown off the crests of waves and streaks on the surface. Given these general changes in the surface, one may expect a qualitatively different behavior in its frictional properties than that suggested by observations in moderate wind conditions". According to observations, winds generate waves with a broad spectrum of wave lengths on the ocean surface. The longest waves hundreds meters long attempt to catch up with the wind, while the steeper intermediately short waves in the range of 0.1-10 *m* break out and play a dominant role in drag formation (Donelan et al., 2004; Soloviev and Lukas, 2006; Chen et al., 2007; Xu et al., 2007). As the wind velocity increases, the sea surface becomes roughened by intermediately short gravity waves, which break after reaching a critical steepness. This breaking produces foam patches and bubbly streaks, and finally the sea becomes almost completely covered by a layer of foam. At wind speeds less than $U_a = 10 m/s$, the foam coverage is negligible, while at wind speeds ~15m/s, foam patches are observed and the coverage-weighed thickness of the foam layer is ≈ 1% (Reul and Chapron, 2003). When the wind speed exceeds the storm force (~24m/s), wave breaking creates streaks of bubbles near the ocean surface. As the wind exceeds the hurricane force (~32m/s), streaks of bubbles combined with patches of foam cover the ocean surface. When the wind speed reaches ~50m/s, a foam layer completely covers the ocean surface (Powell et al., 2003).

The foam development at the air-sea interface provides a possible explanation for the drag reduction phenomenon in strong winds. The principal role of such air-water foam layer in energy dissipation and momentum transfer from hurricane wind to sea waves has been first suggested in (Newell and Zakharov, 1992). In-situ measurements of the foam layer characteristics are scarce (e.g., Reul and Chapron, 2003; Camps et al., 2005 and references therein), and the understanding of its overall role in the physics of a hurricane is still incomplete. A large number of physical parameters determining the phenomenon and lack of respective systematic data additionally complicate the study. Nowadays, there is little hope for comprehensive numerical calculations of the drag coefficient in strong wind conditions that would include a detailed description of wave breaking and foam layer production. Instead, several explanations have been proposed within the atmospheric boundary-layer theory (Powell et al., 2003; Emanuel, 2003; Andreas, 2004; Donelan et al., 2004; Moon et al., 2004; Soloviev and Lukas, 2006; Xu et al., 2007 and references therein), and the corresponding modeling necessarily requires semi-empirical considerations (e.g. Kudryavtsev, 2006).

The present study is carried out using a simple description of the air-water foam, in which its principle features are taken into account – a high density contrast between the foam and surrounding air and water ($\rho_a << \rho_f << \rho_w$), as well as a low liquid content. The study is not concerned with the foam layer formation by the wind-ocean interaction, but rather focuses on how a given foam layer isolates the atmosphere from the ocean. The adopted simplified scheme ignores the effects of wave breaking and air-water mixing, which finally lead to foam production, and postulates instead the existence of a sandwiched foam layer of a finite thickness $L_f$. The value $L_f$ necessary for effective separation of the atmosphere and the ocean is estimated directly from the model. The present model investigates the stability of the corresponding three-fluid system at wind speeds ranging from those of the onset of the effective foam production, $U_a \sim 15\,m/s$, to hurricane winds.

The study is aimed at establishing the stabilization of the sea-water surface by the foam layer in an air-foam-water system due to a shift of unstable intermediately short waves towards smaller scales. Since drag-responsible waves belong to an intermediately short-wavelength part of the spectrum, it is conjectured that their stabilization qualitatively explains the experimentally observed reduction of roughness and, hence, of the drag. Along with stabilizing properties of the foam layer on the water surface, the KHI simultaneously destabilizes the foam-air interface, thus providing a self–sustained mechanism of foam layer formation. Beyond these particular applications, the current work addresses a wide range of three-layer systems with high density contrasts, which are often encountered in geophysics and astrophysics. The stability analysis is carried out by asymptotic expansions both in small air-water density ratio and in water content in foam.

The paper is structured as follows: the dimensionless governing equations of a foam layer sandwiched between air and water are presented in the next Section. Section 3 describes the asymptotic analysis of KHI of the three-layer system and the results of modeling. We discuss and summarize our results in the final section.

## 2. Physical model

*2.1 Background flow*

To demonstrate the stabilization effect of intermediately short drag-responsible waves, a classical KHI problem is considered, which is capable of simulating a broad variety of equilibrium density and wind profiles (Fig. 1). The equilibrium three-layer system of air, foam and water is assumed to be stably stratified, and the wind equilibrium profile is modeled by a piecewise constant function of height with an effective constant free-stream velocity $U_a$ (Fig. 2) equal to the wind speed $U_{10}$ for a reference height $L_{10} = 10m$.

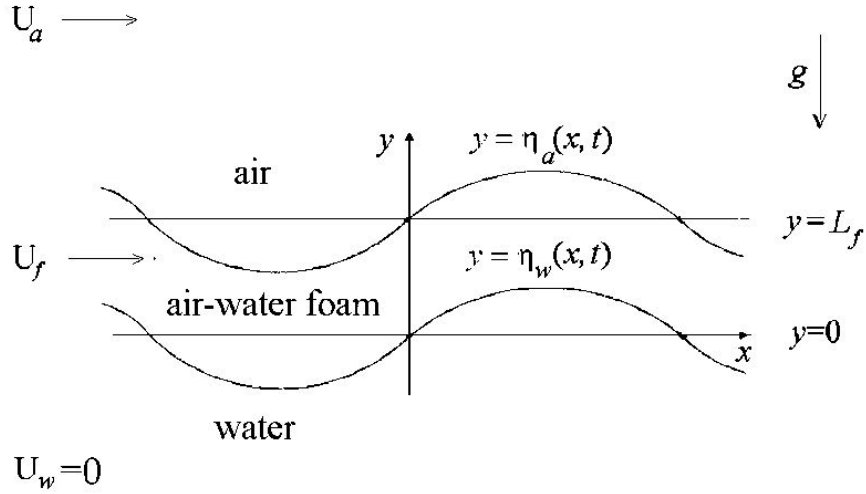

**Figure 1**. Schematic diagram of KHI of a foam layer between the atmosphere and the ocean. $U_a$ and $U_f$ are air and foam velocities; $y = \eta_a(x,t)$ and $y = \eta_w(x,t)$ are air- and water-foam interfaces; $L_f$ is the foam layer thickness; $g$ is the gravity acceleration.

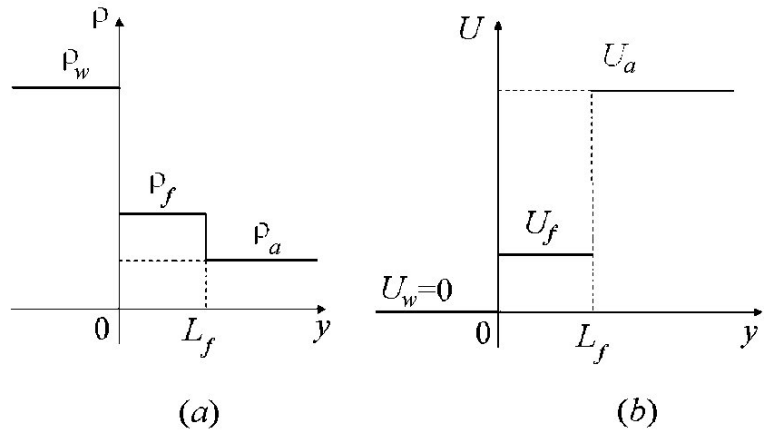

**Figure 2.** Schematic equilibrium profiles of (*a*) density and (*b*) longitudinal velocity vs vertical coordinate in a piecewise constant approximations.

*2.2 Governing relations*

The equations of motion that govern the dynamics of the system are the corresponding Euler equations for incompressible fluids in each of the three layers. Together with interface boundary conditions, these equations yield after some algebra (see Appendix):

$$2(H_a + H_w) + (E-1)(H_a + 1)(H_w + 1) = 0, \tag{1}$$

where

$$H_a(C) = \frac{\rho_a(U_a - C)^2 - (\rho_f - \rho_a)g/k}{\rho_f(U_f - C)^2}, \quad H_w(C) = \frac{\rho_w(U_w - C)^2 - (\rho_w - \rho_f)g/k}{\rho_f(U_f - C)^2}.$$

and $C = \omega/k$ is the phase velocity, $\omega$ is the complex frequency, $k$ is the real wave number.

Before reverting to the general study of the foam layer effect, two limiting cases of foam free ($kL_f$=0) and foam saturated ($kL_f = \infty$) systems are considered. First, it is noted that in the limit of a classical two-fluid system with $kL_f = 0$, or, equivalently, either $\rho_f = \rho_w$ or $\rho_f = \rho_a$, Eq. (1) is reduced to the classical dispersion relation $H_a + H_w = 0$ for KHI (Drazin, 2002):

$$\rho_w k_0 (U_w - C_0)^2 + \rho_a k_0 (U_a - C_0)^2 - (\rho_w - \rho_a)k_0 g = 0, \tag{2}$$

where the subscript 0 denotes foam-free parameters. Second, in the foam-saturated limit $kL_f = \infty$, the quadratic Eq. (1) is decomposed into the following two modes:

$$H_w(C_\infty^{(w)}) + 1 = 0, \quad H_a(C_\infty^{(a)}) + 1 = 0, \tag{3}$$

where each of the modes is described by a quadratic dispersion relation. $C_\infty^{(w)}$ and $C_\infty^{(a)}$ are the eigenvalues corresponding to perturbations of the water-foam and air-foam interfaces (denoted by superscripts $w$ and $a$, respectively, and named, for brevity, water-foam and air-foam modes) in the limit $kL_f = \infty$. Two equations (3) are transformed into the classic Eq. (2) by the variables $(\rho_f, U_f)$ substitution with either $(\rho_a, U_a)$ or $(\rho_w, U_w)$, respectively. This demonstrates that the two modes in Eqs. (3) are induced by KHI of the water-foam and air-foam interfaces, respectively.

Note that the complete dispersion relation (1) can be easily solved numerically for any known foam layer thickness $L_f$ and velocity $U_f$. However, since the foam layer parameters are badly known values in strong wind conditions, below dispersion relation (1) is solved asymptotically, which allows to evaluate the values of $L_f$ and $U_f$ using the conditions of effective separation of the atmosphere from the ocean by a foam layer, as well as experimental data for roughness and dimensional considerations.

## 3. Asymptotic analysis

The stability analysis is carried out by asymptotic expansions in two small parameters: air-water density ratio and water content in foam.

*3.1 Classic air-water system*

First, the limit of low air-water density ratio $\rho_a/\rho_w = \varepsilon^2 \ll 1$ ($\varepsilon^2 \approx 10^{-3}$) is applied to the classical two-layer case described by Eq.(2) with $U_w = 0$ in order to estimate the orders in $\varepsilon$ of the system parameters, which strongly influence the system stability:

$$\omega_0^{(w)} = \sqrt{k_0 g - \varepsilon^2 k_0^2 U_a^2} + O(\varepsilon^2 k_0 U_a, \varepsilon g k_0/\omega_0). \tag{4}$$

Doing so, it can be concluded that the classical two-fluid KHI is excited in the short wavelength regime:

$$k_0 L_* \sim k_0^* L_* = \varepsilon^{-2}, \quad \omega_0^{(w)} L_*/U_* \sim \varepsilon^{-1}, \quad C_0^{(w)}/U_* \sim \varepsilon,$$

$$k_0^* = 1/(\varepsilon^2 L_*) \quad L_* = U_a^2/g, \quad U_* = U_a. \tag{5}$$

Here the subscript and superscript asterisks denote characteristic scales and marginal values of the parameters, respectively.

Reverting to the general air-foam-water systems, it is assumed according to a characteristic feature of foams that water content within the foam, $\alpha_w$ ($\alpha_w \sim 0.05$), is low. As a result, $\alpha_w$ is scaled with $\varepsilon$ and yields

$$\frac{\rho_f}{\rho_*} \approx \alpha_w \sim \varepsilon, \quad \frac{\rho_a}{\rho_*} = \varepsilon^2, \tag{6}$$

where $\rho_* = \rho_w$, $\rho_f = \alpha_w \rho_w + \alpha_a \rho_a$ is the foam density; $\alpha_w, \alpha_a$ are water and air volume fractions within the foam; $\alpha_w + \alpha_a = 1$.

Although the eigenvalues for water- and air-foam modes are coupled by the complete dispersion relation (1), they have different orders in $\varepsilon$ and can be evaluated separately.

*3.2 Water-foam mode*

Assuming now that the three-fluid system operates in the same regime as that giving rise to the KHI in the classic two-fluid system, the scales (5) are adopted

$$kL_* \sim \frac{1}{\varepsilon^2}, \quad \frac{\omega^{(w)} L_*}{U_*} \sim \frac{1}{\varepsilon}, \quad \frac{C^{(w)}}{U_*} \sim \varepsilon. \tag{7}$$

As shown below, estimates (7) select the water-foam mode of the entire dispersion relation (1). Following the scaling (5) and (7), the wave number and frequency are rescaled as follows:

$$\hat{k} = \frac{k}{k_0^*} \sim \varepsilon^0, \quad \hat{\omega}^{(w)} = \frac{\omega^{(w)}}{\sqrt{gk_0^*}} \equiv \hat{C}^{(w)}\hat{k} \sim \varepsilon^0, \quad \hat{C}^{(w)} = \frac{C}{\sqrt{k_0^*/g}} \sim \varepsilon^0. \tag{8}$$

Further, assuming for the water-foam mode that the foam layer thickness is much less than the characteristic length, $L_f / L_* \ll 1$ ($L_* = U_a^2 / g \sim 20m$ for $U_a \sim 15 m/s$, and $L_* \sim 160m$ for $U_a \sim 40 m/s$), while the foam velocity is much less then the wind velocity and much larger than the phase velocity $\varepsilon \sim C^{(w)} / U_* \ll U_f / U_* \ll 1$:

$$U_f / U_* \sim \varepsilon^a, \quad L_f / L_* \sim \varepsilon^b, \quad 0<a<1, \; 0<b. \tag{9}$$

This yields the following estimates for the dispersion relation Eq. (1)

$$H_a \sim H_w \sim \varepsilon^{1-2a}, \quad E \sim \exp(\varepsilon^{b-2}). \tag{10}$$

Inserting the scaling (10) into Eq. (1) and applying the principle of the least degeneracy of the problem (Van Dyke, 1964) to this equation, results in $a = 1/2$, $b = 2$, which means that:

$$\frac{U_f}{U_*} \sim \varepsilon^{1/2}, \quad \frac{L_f}{L_*} \sim \frac{\lambda_0^*}{L_*} \sim \frac{1}{k_0^* L_*} \sim \frac{\rho_a}{\rho_*} = \varepsilon^2, \tag{11}$$

where $\lambda_0^* = 2\pi / k_0^*$.

Then the dispersion relation (1) yields for the water-foam mode, to leading order in $\varepsilon$

$$\hat{\omega}^{(w)} = \sqrt{\frac{2(\hat{k} - \hat{k}^2) - (E-1)(\hat{k}^2 K_f - \hat{k})(K_f^{-1} + 1)}{2 + (E-1)(K_f^{-1} + 1)}}, \quad E = \exp(2\hat{k}\hat{L}_f). \tag{12}$$

Here $\hat{\omega}^{(w)}$, $\hat{C}^{(w)}$, $\hat{k}$ and $\hat{L}_f$ denote the values rescaled to the order $\varepsilon^0$ of the frequency, phase velocity, wave number and the foam thickness, while $K_f$ ($0<K_f<1$) is the ratio of the foam-to-air momentum flux:

$$\hat{\omega}^{(w)} = \frac{\omega^{(w)}}{\sqrt{g k_0^*}} \equiv \hat{k} \hat{C}^{(w)}, \quad \hat{C}^{(w)} = \frac{C^{(w)}}{\varepsilon U_*}, \quad \hat{k} = \frac{k}{k_0^*}, \quad \hat{L}_f = k_0^* L_f, \quad K_f = \frac{\rho_f U_f^2}{\rho_a U_a^2}. \tag{13}$$

Thus, the system stability is parameterized by the dimensionless foam thickness and velocity or, equivalently, $\hat{L}_f = k_0^* L_f$ and $K_f$, where $K_f$ is the momentum flux ratio, and the dimensionless foam thickness $k_0^* L_f$ has the meaning of bulk foam Richardson number $Ri_f$ scaled by $\rho_a / \rho_f \approx \varepsilon^2 / \alpha_w \sim \varepsilon$:

$$\hat{Ri}_f = k_0^* L_f \sim \varepsilon^0, \quad (Ri_f = -g \frac{\Delta \rho}{\rho_f} \frac{L_f}{(\Delta U)^2} \approx \frac{\rho_a}{\rho_f} \hat{Ri}_f), \quad K_f = \frac{\rho_f U_f^2}{\varepsilon^2 \rho_w U_*^2} \sim \varepsilon^0. \tag{14}$$

Here $Ri_f$ describes the competition between the opposing shear and buoyancy effects; $\Delta U = U_a - U_w \equiv U_*$ and $\Delta \rho = \rho_a - \rho_w = -\rho_*(1 - \varepsilon^2)$ are the jumps of the velocity and density of the foam layer; the reduced gravity $g\Delta\rho/\rho_f$ is the vertical gravity acceleration $g$ factored by the density step $\Delta\rho$ and made dimensionless by the density within the foam layer $\rho_f$.

Two particular limits of the dispersion relation (12) for the water-foam mode are readily obtained for small $\varepsilon$, namely the foam-free ($H_a + H_w = 0$ for $L_f = 0$) and foam-saturated ($H_w + 1 = 0$ for $L_f = \infty$) limits, respectively:

$$\hat{\omega}_0^{(w)} \equiv \frac{\omega_0^{(w)}}{\sqrt{gk_0^*}} = i\sqrt{\frac{k^2}{k_0^{*2}} - \frac{k}{k_0^*}}, \quad k_0^* L_f = 0, \tag{15}$$

$$\hat{\omega}_\infty^{(w)} \equiv \frac{\omega_\infty^{(w)}}{\sqrt{gk_\infty^*}} = i\sqrt{\frac{k^2}{k_\infty^{*2}} - \frac{k}{k_\infty^*}}, \quad k_0^* L_f = \infty, \tag{16}$$

where

$$k_\infty^* = k_0^* / K_f \quad (0 < K_f < 1). \tag{17}$$

Equation (16) differs from Eq. (15) by $k_0^*$, $\omega_0^{(w)}$ substitution with $k_\infty^*$, $\omega_\infty^{(w)}$. The comparison of two limits (15) and (16) demonstrates the stabilizing effect of the foam layer due to the decrease in the marginal wavelength from the foam free $\lambda_0^* = 2\pi/k_0^*$ to the foam-saturated $\lambda_\infty^* = 2\pi/k_\infty^*$ value. The growth rate $\omega_i$ decreases from the foam-free $\omega_{i0}$ to the foam-saturated $\omega_{i\infty}$ value.

Relation (17) resolved with respect to $K_f$ allows to express it or, alternatively, $U_f$ through the ratio of the foam-saturated-to-foam-free wavelengths $\lambda_\infty^*/\lambda_0^*$ ($U_f$, in addition, depends on the volume content of water $\alpha_w$)

$$K_f = \frac{\lambda_\infty^*}{\lambda_0^*} \equiv \frac{\lambda_\infty^*}{2\pi\varepsilon^2 L_*} \sim \varepsilon^0, \quad \frac{U_f}{\sqrt{gL_*}} = \sqrt{\frac{\lambda_\infty^*}{L_* 2\pi\alpha_w}} \equiv \varepsilon \frac{U_a}{\sqrt{gL_*}} \sqrt{K_f/\alpha_w} \sim \varepsilon^{1/2}. \tag{18}$$

First, an estimation from above for $K_f$ can be obtained assuming that the foam-saturated limit has been achieved and $\lambda \approx \lambda_\infty^*$. Since $0 < K_f < 1$, Eqs. (18) for $K_f$ also yields the upper bound for $\lambda_\infty^*$:

$$\lambda_\infty^* < \Lambda^* = 2\pi\varepsilon^2 L_* \tag{19}$$

(see estimation for $K_f$ based on experimental data and dimensional grounds in Section 4).

We assume that intermediately-short waves under consideration belong to the wavelength interval of drag-responsible waves $\sim 0.1-10\ m$ (Chen et al., 2007), and wind speeds – to the interval of foam-generated winds between $U_a = 12\ m/s$ corresponding to the onset of foam generation (Reul and Chapron, 2003) and $U_a \approx 50\ m/s$ corresponding to a complete coverage of the ocean surface. Then relation (19) yields $\Lambda^* \sim 0.15 m$ for $U_a \sim 15 m/s$ and $\Lambda^* \sim 1\ m$ for $U_a \sim 40 m/s$. Choosing for further estimations the intermediate value of $K_f \approx 0.5$, $0 < K_f < 1$ (see discussion in Section 4), a typical value of $\alpha_w \sim 0.05$, and using Eq. (18) resolved with respect to $\lambda_\infty^*$ yields: $U_f = \varepsilon U_a \sqrt{K_f/\alpha_w} \sim 1.5 m/s$, $\lambda_\infty^* = 2\pi\varepsilon^2 K_f U_*^2/g \sim 0.1 m$ for $U_a \sim 15 m/s$, and $U_f \sim 4 m/s$, $\lambda_\infty^* \sim 1 m$ for $U_a \sim 40 m/s$.

Figure 3 depicts a normalized growth rate $\hat\omega_i^{(w)}$ vs. wavenumber $\hat k$ for a typical foam-layer thickness $\hat{Ri}_f \equiv \hat L_f = k_0^* L_f$ and a fixed ratio of the foam-to-air momentum flux $K_f = k_0^*/k_\infty^* = 0.5$. The growth rate decreases as the foam thickness is increased and approaches its foam-saturated limit $k_0^* L_f = \infty$ (given by Eq. (16)) as early as at the effective value of the foam layer thickness: $\hat{Ri}_f = k_0^* L_f \approx 1$, and further increase of $L_f$ is ineffective. This allows us to evaluate the foam layer thickness necessary for effective separation of the atmosphere and the ocean as follows: $\hat L_f \approx 1$, or equivalently, $L_f^{(ef)} \approx \varepsilon^2 L_*$. The corresponding marginal value of wavenumber (wavelength) that bounds from above (below) the region of stability may be found as an intersection point of the effective foam-saturated curve $\hat L_f = 1$ and abscissa axis $\hat k_\infty^* = 2$ ($\hat\lambda_\infty^* = 2\pi/\hat k_\infty^* = \pi$) in Fig. 3. Figure 3 demonstrates also the instability shift towards smaller wavelength scales. Finally, note that the dimensional effective thickness of the foam layer $L_f^{(ef)} = \varepsilon^2 U_a^2/g$ is larger for stronger winds.

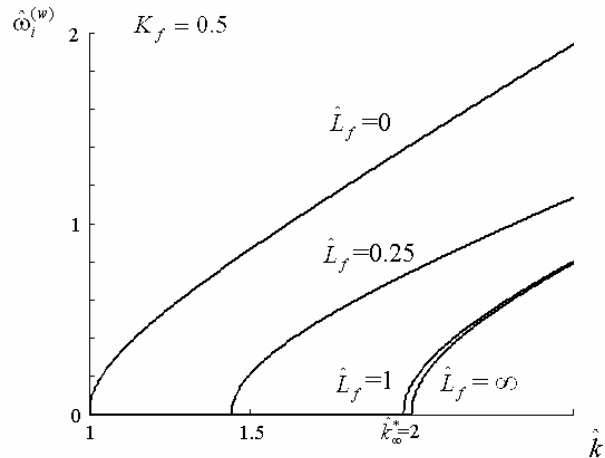

**Figure 3**. Growth rate $\hat\omega_i^{(w)} = \omega_i^{(w)}/\sqrt{gk_0^*}$ vs. wave number $\hat k = k/k_0^*$ for the water-foam mode.

Typical foam-layer thickness, $\hat{R}i_f = \hat{L}_f \equiv k_0^* L_f = 0; 0.25; 1.0, \infty$, and foam-to-air momentum flux ratio

$$K_f = k_0^* / k_\infty^* = 0.5.$$

The dependence of normalized growth rate $\hat{\omega}_i^{(w)}$ on the foam-layer thicknesses $\hat{R}i_f = \hat{L}_f \equiv k_0^* L_f$ and a fixed $K_f = k_0^* / k_\infty^* = 0.5$ is depicted in Fig. 4. For sufficiently short waves ($k/k_0^* > 1/K_f$), $\hat{\omega}_i$ strongly drops from the foam-free value $k_0^* L_f = 0$ to its effective saturation level at foam layer thickness $k_0^* L_f \approx 1$. The growth rates of perturbations with longer waves ($k/k_0^* < 1/K_f$) sharply decrease with $k_0^* L_f$ increasing from zero until an effective stabilization at a finite value of $k_0^* L_f^{(ef)} \approx 1$ is achieved. These two cases are separated by a threshold curve ($k/k_0^* = 1/K_f$) for which the growth rate $\hat{\omega}_i$ vanishes at $k_0^* L_f \gg 1$.

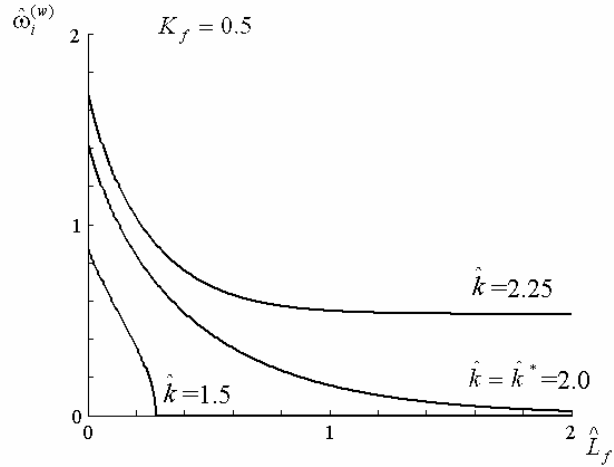

**Figure 4.** Growth rate $\hat{\omega}_i^{(w)}$ vs. foam-layer thickness $\hat{R}i_f = \hat{L}_f \equiv k_0^* L_f$ for water-foam mode at typical wave numbers $\hat{k} = k/k_0^*$ and ratio of the foam-to-air momentum flux $K_f = k_0^* / k_\infty^* = 0.5$.

The marginal wavenumber $k^*$ satisfies the eigenvalue equation for the three-layer system ($\omega_i^{(w)} = 0$):

$$\exp(2k^* L_f) = 1 - \frac{2}{1 + K_f^{-1}} \frac{1 - k^*/k_0^*}{1 - k^*/k_\infty^*}. \tag{20}$$

In Fig. 5 nontrivial solutions of Eq. (20) $k^*/k_0^*$ ($1 \le k^*/k_0^* < K_f^{-1}$, $0 < K_f < 1$) are depicted vs $k_0^* L_f$ for several values of $K_f$. The value of $K_f$ is estimated on the basis of experimental data and scaling arguments (see Section 4 below).

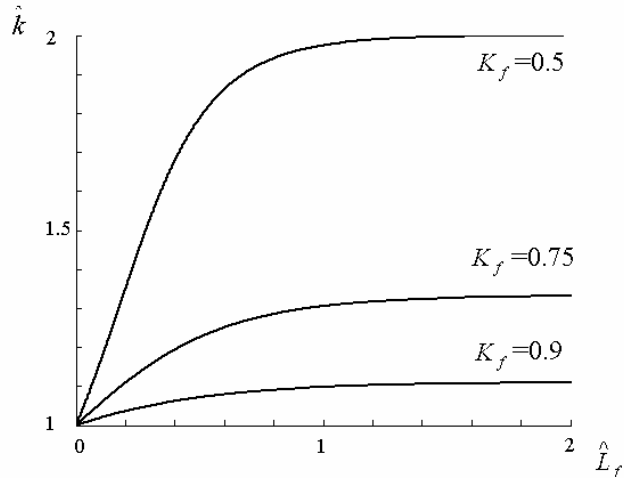

**Figure 5**. Marginal wave number $\hat{k}^* = k^*/k_0^*$ vs foam layer thickness $\hat{R}i_f = \hat{L}_f \equiv k_0^* L_f$ for water-foam mode at the momentum flux ratio $K_f = 0.5, 0.75, 0.9$.

*3.3 Air-foam mode and inputs of air/water-foam modes into the water/air-foam interfaces*

Now the air-foam mode of Eq. (1) is considered assuming that the orders of the foam velocity and thickness are given by estimations (11) and replacing the scaling (18) of the water-foam mode by that of the air-foam mode, which is found as follows:

$$kL_* \sim k_0^* L_* = \frac{1}{\varepsilon^2}, \quad \frac{\omega^{(a)} L_*}{U_*} \sim \frac{1}{\varepsilon^{3/2}}, \quad \frac{C^{(a)}}{U_*} \sim \varepsilon^{1/2}. \tag{21}$$

Then Eq. (1) in leading order in $\varepsilon$ is reduced to

$$H_a + 1 = -2/(E-1), \qquad E = \exp(2\hat{k}\hat{L}_f). \tag{22}$$

In the foam-saturated limit ($\hat{L}_f = \infty$), Eq. (22) evidently coincides with the air-foam mode in the second of Eqs. (3). Hence, to the leading order in $\varepsilon$, the unstable air-foam mode is

$$\hat{\omega}^{(a)} = \hat{k}(\hat{U}_f \pm i\sqrt{\frac{1}{\hat{\rho}_f}\frac{E-1}{E+1}}). \tag{23}$$

Here rescaled values of the order of $\sim \varepsilon^0$ are introduced:

$$\hat{\omega}^{(a)} = \frac{\omega^{(a)}\sqrt{\varepsilon}}{\sqrt{gk_0^*}} \equiv \hat{k}\hat{C}^{(a)}, \quad \hat{C}^{(a)} = \frac{C^{(a)}}{U_*\sqrt{\varepsilon}}, \quad \hat{k} = \frac{k}{k_0^*}, \quad \hat{U}_f = \frac{U_f}{U_*\sqrt{\varepsilon}}, \quad \hat{\rho}_f = \frac{1}{\varepsilon}\frac{\rho_f}{\rho_w} \approx \frac{\alpha_w}{\varepsilon}.$$

Phase velocity of the air-foam mode equals the foam layer velocity, while the growth rate increases from 0 in the foam-free limit at $k_0^* L_f = 0$ to the maximal foam-saturated value $\hat{k}/\sqrt{\hat{\rho}_f}$ at $k_0^* L_f = \infty$ (Figs. 6 and 7). Inputs of the air-foam mode into the eigenfunctions of the air- and foam-occupied domains vanish in $kL_f$ and approach the foam-saturated limit when the foam thickness exceeds the effective value $L_f \approx L_f^{(ef)}$ ($k_0^* L_f^{(ef)} \approx 1$).

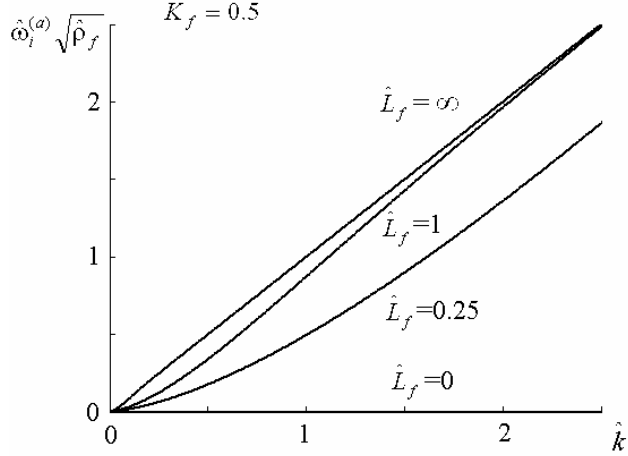

**Figure 6.** Growth rate $\hat{\omega}_i^{(a)} \sqrt{\hat{\rho}_f} = \omega_i^{(a)} \sqrt{\rho_f / \rho_w} / \sqrt{gk_0^*}$ vs wave number $\hat{k} = k/k_0^*$ for the air-foam mode at a typical foam-layer thickness $\hat{Ri}_f = \hat{L}_f \equiv k_0^* L_f$ and foam-to-air momentum flux ratio

$$K_f = k_0^* / k_\infty^* = 0.5.$$

Now the inputs of the air-foam mode into the water-foam interface, as well as the inputs of the water-foam mode into the air-foam interface can be calculated. The eigenfunctions of the water-foam and air-foam modes normalized to the order of $\varepsilon^0$ are as follows:

$$\hat{U}_a = \frac{\tilde{U}_a}{U_*}, \ \hat{V}_a = \frac{\tilde{V}_a}{U_*}, \ \hat{P}_a = \frac{\tilde{P}_a}{\varepsilon^2 \rho_* U_*^2}, \ \hat{U}_w = \frac{\tilde{U}_w}{\varepsilon U_*}, \ \hat{V}_w = \frac{\tilde{V}_w}{\varepsilon U_*}, \ \hat{P}_w = \frac{\tilde{P}_w}{\varepsilon^2 \rho_* U_*^2},$$

$$\hat{\eta}_a = \hat{k} k_0^* \sqrt{E} \tilde{\eta}_a, \ \hat{\eta}_w = \hat{k} k_0^* \sqrt{E} \tilde{\eta}_w, \ \hat{U}_f = \frac{\tilde{U}_f}{\sqrt{\varepsilon} U_*}, \ \hat{V}_f = \frac{\tilde{V}_f}{\sqrt{\varepsilon} U_*}, \ \hat{P}_f = \frac{\tilde{P}_f}{\varepsilon^2 \rho_* U_*^2}. \quad (24)$$

Using Eqs. (24) and substituting the eigenvalues for the water-foam and air-foam modes $\hat{C} = \hat{C}^{(w)}$ and $\hat{C} = \hat{C}^{(a)}$ from Eqs. (12) and (23), respectively, into relations (A9) and (A12), yields for the eigenfunctions of the air-foam and water-foam modes:

$$\frac{\hat{\eta}_w^{(a)}}{\hat{\eta}_a^{(a)}} = \frac{\hat{U}_w^{(a)}}{\hat{\eta}_a^{(a)}} = \frac{\hat{V}_w^{(a)}}{\hat{\eta}_a^{(a)}} = \frac{\hat{P}_w^{(a)}}{\hat{\eta}_a^{(a)}} = 0, \quad \frac{\hat{U}_a^{(a)}}{\hat{\eta}_a^{(a)}} = -i \frac{\hat{V}_a^{(a)}}{\hat{\eta}_a^{(a)}} = -\frac{\hat{P}_a^{(a)}}{\hat{\eta}_a^{(a)}} = 1,$$

$$-i \frac{\hat{U}_{+f}^{(a)}}{\hat{\eta}_a^{(a)}} = i \frac{\hat{U}_{-f}^{(a)}}{\hat{\eta}_a^{(a)}} = \frac{\hat{V}_{+f}^{(a)}}{\hat{\eta}_a^{(a)}} = -\frac{\hat{V}_{-f}^{(a)}}{\hat{\eta}_a^{(a)}} = \pm \sqrt{\frac{1}{\hat{\rho}_f} \frac{1}{E^2-1}}, \quad \frac{\hat{P}_{\pm f}^{(a)}}{\hat{\eta}_a^{(a)}} = -\frac{1}{E+1}, \quad \text{(air-foam mode)} \quad (25)$$

$$\frac{\hat{\eta}_a^{(w)}}{\hat{\eta}_w^{(w)}} = \frac{2K_f}{(1+K_f)\sqrt{E} - (1-K_f)/\sqrt{E}}, \quad \frac{\hat{U}_w^{(w)}}{\hat{C}^{(w)} \hat{\eta}_w^{(w)}} = i \frac{\hat{V}_w^{(w)}}{\hat{C}^{(w)} \hat{\eta}_w^{(w)}} = \frac{\hat{P}_w^{(w)}}{\hat{C}^{(w)2} \hat{\eta}_w^{(w)}} = \frac{1}{\sqrt{E}},$$

$$\frac{\hat{U}_a^{(w)}}{\hat{\eta}_a^{(w)}} = -i\frac{\hat{V}_a^{(w)}}{\hat{\eta}_a^{(w)}} = -\frac{\hat{P}_a^{(w)}}{\hat{\eta}_a^{(w)}} = 1, \quad -i\frac{\hat{V}_{-f}^{(w)}}{\hat{\eta}_a^{(w)}} = \frac{\hat{U}_{-f}^{(w)}}{\hat{\eta}_a^{(w)}} = \frac{1+K_f}{2\sqrt{K_f\hat{\rho}_f}}, \quad \frac{\hat{P}_{-f}^{(w)}}{\hat{\eta}_a^{(w)}} = -\frac{1+K_f}{2},$$

$$i\frac{\hat{V}_{+f}^{(w)}}{\hat{\eta}_a^{(w)}} = \frac{\hat{U}_{+f}^{(w)}}{\hat{\eta}_a^{(w)}} = \frac{1}{2E}\frac{1-K_f}{\sqrt{K_f\hat{\rho}_f}}, \quad \frac{\hat{P}_{+f}^{(w)}}{\hat{\eta}_a^{(w)}} \approx -\frac{1-K_f}{2E}. \qquad \text{(\textit{water-foam mode})} \quad (26)$$

Here one of the arbitrary magnitudes for each of the two modes can be set equal to unity, for instance, $\hat{\eta}_a^{(a)} = \hat{\eta}_w^{(w)} = 1$. In particular, the former of Eqs. (25) demonstrates zero input of the air-foam mode into the water-foam interface. Hence the air-foam mode does not perturb the water-foam interface, and stability of the water-foam interface is completely described by the water-foam mode. Asymptotically in high $kL_f$ the same is valid for the input of the water-foam mode into the air-foam interface, which approaches the foam-saturated limit when the foam thickness exceeds the effective value $L_f^{(ef)}$. Thus, duality of the KHI of the three-fluid (air-foam-water) configuration takes place, which demonstrates a reduction of the water-surface instability compared with that in the two-fluid (air-water) system, and simultaneously exhibits a instability of the air-foam interface.

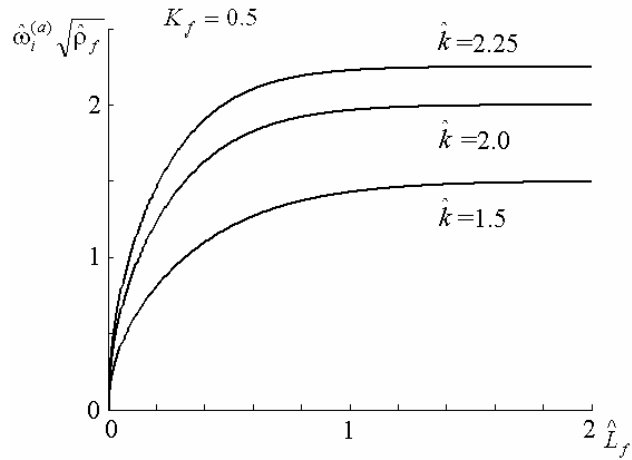

**Figure 7**. Growth rate $\hat{\omega}_i^{(a)}\sqrt{\hat{\rho}_f} = \omega_i^{(a)}\sqrt{\rho_f/\rho_w}/\sqrt{gk_0^*}$ vs foam-layer thickness $\hat{Ri}_f = \hat{L}_f \equiv k_0^* L_f$ for the air-foam mode at a typical wave number $\hat{k} = k/k_0^*$ and ratio of foam-to-air momentum flux. $K_f = k_0^*/k_\infty^* = 0.5$.

## 4. Summary, discussion and conclusions

The present model postulates drag reduction in a non-equilibrium three-fluid configuration rather than calculates the drag value. The stability analysis above is aimed at the explanation of possible physical mechanisms which support the postulated drag reduction. As mentioned above, the adopted simplified scheme ignores the effects of wave breaking and air-water turbulent mixing, which finally lead to foam production. Instead, the model postulates the existence of a foam layer with averaged finite thickness, $L_f$ and velocity $U_f$ sandwiched between the water at rest and moving air. The value of $L_f$ is evaluated directly from the model basing on the results of stability analysis as the foam thickness necessary for effective separation of the atmosphere and the ocean. Since in real conditions the foam layer is unsteady and inhomogeneous, it is conjectured in the present model that the input and output characteristics of the equilibrium thickness of the foam layer are averaged over the coating area. The dimensionless foam layer velocity $U_f$ or, equivalently, the momentum flux ratio $K_f = \rho_f U_f^2 / \rho_a U_a^2$, are estimated using experimental data for the observed drag reduction coefficient. In this model of the sandwiched system the water is at rest, and the resulting dimensionless foam velocity is found to be asymptotically small ($U_f / U_a \sim \sqrt{\varepsilon}$).

The analysis is performed for high density contrast systems, for which the traditional oceanographic approach of using Bussinesq approximation is not applicable. The analysis of the KHI in a three-layer system with high density contrasts is treated asymptotically in two small parameters: air-water density ratio $\sim \varepsilon^2$ and water content in the foam $\alpha_w \sim \varepsilon$. The system stability is parameterized by the dimensionless foam velocity $U_f$ and thickness $L_f$ (or, equivalently, the dimensionless momentum flux ratio $K_f$ and Richardson number $\hat{R}i_f$). Due to the lack of observations or modeling data in strong wind conditions, they are first estimated asymptotically as $L_f/L_g \sim \varepsilon^2$, $U_f/U_a = \varepsilon\sqrt{K_f/\alpha_w} \sim \sqrt{\varepsilon}$ ($0 < K_f < 1$), by applying the asymptotic principle of the least degeneracy of the problem. Such values of parameters $K_f$ and $\hat{R}i_f$ correspond to the strongest influence of the foam layer on the system stability. Then the value of $L_f^{(ef)}$ necessary for effective separation of the atmosphere and the ocean is evaluated using the condition that the growth rate approaches the foam- saturated value, if $\hat{L}_f^{(ef)} = 1$ ($L_f^{(ef)} = L_g \varepsilon^2$, $L_g = U_a^2/g$), and the further increase of $L_f$ is ineffective, as if the foam layer is of infinite thickness. This yields an evaluation for $L_f^{(ef)} = \varepsilon^2 L_g$ and leaves a single free parameter $U_f$ (or $K_f$) that parameterizes the system. Choosing the intermediate value of $K_f \approx 0.5$ (see below) for our estimations, we can estimate the foam-layer velocity and thickness, as well as the corresponding wavelength. The present modeling provides also the wavelengths corresponding to lower bounds of the stability region. This yields for (i) $U_a \sim 15 m/s$ and (ii) $U_a \sim 40 m/s$, respectively ($\alpha_w \sim 0.05$): (i) $L_f^{(ef)} \sim 0.02 m$, $U_f \sim 1.5 m/s$, $\lambda_\infty^* \sim 0.1 m$; (ii) $L_f^{(ef)} \sim 0.16 m$, $U_f \sim 4 m/s$, $\lambda_\infty^* \sim 1 m$. These values of $L_f^{(ef)}$ agree by the order of magnitudes with the observation data (Reul and Chapron, 2003). According to these data for foams generated by breaking ocean waves, an increase in the wind speed from 7 to 20 m/s corresponds to a coverage-weighed foam layer thickening by about 1 cm and 3.5 cm, respectively.

The KHI of the three-fluid (air-foam-water) configuration demonstrates a reduction of the water-surface instability compared with that of the two-fluid (air-water) system, and simultaneously exhibits an instability of the air-foam interface. The established duality of the KHI allows us to conjecture that the KHI provides a self-sustaining mechanism for the three-fluid system existence due to simultaneous destruction and regeneration of the foam layer. Thus, in the stable part of the spectra of the foam-water mode, foam destruction due to the instability of the air-foam mode returns the water surface to the unstable state of the foam-free system, and further wave breaking and foam regeneration should lead to the system stabilization, etc.

The present modeling exhibits the instability shift towards smaller wavelength scales. It demonstrates a new effective mechanism of the water surface stabilization by a foam layer, which effectively separates the atmosphere from the ocean due to high density contrasts in the three-fluid system. Since drag-responsible waves belong to the intermediately short part of the spectrum, it is conjectured that their stabilization qualitatively explains the experimentally observed reduction of the roughness and, hence, of the drag. The results are physically transparent, since in the foam-saturated system the foam layer separates the air flow from the sea surface, and the three-fluid system becomes close to a two-fluid foam-water system. Formally, this corresponds to the substitution of the foam density and velocity instead of those parameters for the air in the classic two-fluid model. It can be supported by the following dimensionality arguments. Since for $K_f = 0.5$ the foam layer reduces the foam-free wavelength $\lambda_0^*$ approximately by a factor of $1/K_f = 2$ to the foam-saturated limit $\lambda_\infty^*$ at $L_f \approx L_f^{(ef)} = \varepsilon^2 L_g$. This scale-down in the characteristic unstable length scales provides a qualitative link between the linear stability modeling and the role of the foam layer in the air-sea momentum exchange. To show that, a widely accepted correlation between the ocean surface roughness $z/\lambda$ and wave steepness $h/\lambda$ (Taylor and Yelland, 2001) is adopted here for their local values

$z/\lambda = F(h/\lambda).$  (27)

Noting that the breaking process does not completely destroy the waves, but rather tears off their tops, when their steepness exceeds a critical value (which is determined by nonlinear effects), it is suggested that the foam production occurs when the critical steepness 1/10 (Fringer and Street, 2003) is achieved for drag-responsible breaking waves. Such simple scaling for the roughness-to-wavelength ratio provides estimation for the numerical value $K_f$ from the observations data for roughness. Substituting correlation (27) into the definition (18) of $K_f$ yields

$$K_f = \lambda_\infty^* / \lambda_0^* \approx z_\infty / z_0,$$  (28)

where $\lambda_\infty^* / \lambda_0^*$ is the ratio of the foam-saturated-to-foam-free wavelengths, and $z_\infty / z_0$ is the ratio of the foam-saturated roughness to the foam-free roughness. Taking as estimations for (i) $z_\infty$ and (ii) $z_0$ in (28) the roughness values (averaged within admissible errors) (i) measured in natural conditions by Powell et al. (2003) and (ii) correlations for roughness obtained by Large and Pond (1981) by extrapolation from low-to-high wind conditions with no account of the foam layer effect (see both dependences in Fig. 3b in Powell et al., 2003), yields roughly $K_f \sim 0.5$ in the range of wind speed $25 < U_a < 40$ m/s ($U_a \sim U_{10}$). As a result, the foam layer reduces both the roughness and the wave length by approximately a factor of $1/K_f = 2$ as compared with the results of extrapolation from low- to high-wind conditions. Additionally, drag reduction takes place due to a qualitative similarity in the behavior of roughness and drag (Powell et al., 2003).

Concerning the general mechanism governing the stability properties of wind waves, note that Kelvin-Helmholtz and Miles theories describe two principal mechanisms of the energy and momentum transfer from wind to waves in a foam-free system. The KHI theory considers a tangential discontinuity between uniform flows of air and water, while the quasi-laminar Miles model takes turbulence effects into account through the unperturbed wind profile, which continuously varies with the height above the interface. The Miles mode correctly predicts the minimum wind speed that gives rise to interface instability, while the KH mode overestimates it. On this basis, a commonly accepted opinion widespread in geophysics and hydrodynamics (see Barnett et al., 1975, Kraus and Businger 1994 and a brief review of Shtemler et al., 2008) is that the Miles model is an improvement of the KH model, and the KH regime has no physical meaning in the context of the surface wave generation. In fact, these two mechanisms of the instability operate in quite different scales. The Miles mode dominates at low wind speed, in particular, it predicts the minimum wind speed that gives rise to water surface instability, while the generalized KH regime dominates for strong winds and raises intermediately-short waves much longer than the Miles mode (Morland and Saffman, 1993; Caponi et al., 1992; Shtemler et al., 2008), while for very strong winds the generalized KH mode is reduced to the classic KHI of tangential discontinuity between two uniform flows (Shtemler et al., 2008). Since the foam layer thickness and velocity are badly known in high wind conditions, the piecewise constant approximation (describing the classic KHI in two-fluid limit) is applied for the foam and air velocities, both for simplicity and for the possibility to qualitatively explain major properties of the water surface stabilization. It is believed that the stabilization of water surface established here for a piecewise constant wind profile reflects the general property of intermediately short waves for more realistic wind profiles.

Finally, let us discuss the arguments against the KHI as a cause of the sea-wave generation in Kraus and Businger (1994): "There are two reasons why Kelvin-Helmholtz instability can not be primary cause of wave generation on the sea surface". First, as was already mentioned, it overestimates the minimal wind velocity at which waves could grow. This argument can be also removed in the context of the present study, since it concerns the foam layer produced at wind velocities much larger than critical values (both experimentally observed critical wind and theoretical threshold corresponding to KHI). "Second, air pressure perturbations in phase or antiphase with surface elevation don not transfer the moment to the water". In this connection note that although presence of the foam layer between the atmosphere and the ocean (postulated in the present study, see also Shtemler et al. 2007) leaves description of its production out of consideration, this assumes the momentum flux transfer within the equilibrium system. Thus, parameter $0<K_f<1$, defined in Eq. (14), is nothing else as the momentum flux ratio at the air-foam interface, and the value $1/K_f=2$ chosen above for modeling corresponds to experimentally observed about two times reduction of drag.

Although in strong wind conditions, the bubbly liquid, spray and foam coexist within a layer that separates the atmosphere and the ocean, the present model assumes the foam alone within the layer. Indeed, gas-liquid foams strongly differ from other two-phase mixtures composed from the same constituents. Natural sea foams are formed of bubbles separated by thin liquid gaps, while bubbly-liquids/sprays consist of bubbles/drops surrounded by thick liquid/gas layers. Large density contrast in the air-foam-water system ($\rho_a << \rho_f \sim \alpha_w \rho_w << \rho_w$) follows from another characteristic feature of foam – low water content ($\alpha_w \sim 0.05$) within them. In turn, a three-fluid system with a foam layer of large density contrasts is qualitatively distinguished from those with layers of bubbly liquid or spray. Bubbly liquids and sprays (with the densities $\rho_b$ and $\rho_s$), are hard to distinguish by density values from water and air, respectively, since $\rho_s \sim \rho_a \ll \rho_f \ll \rho_b \sim \rho_w$, because typically $\rho_b = \alpha_w \rho_w + \alpha_a \rho_a \approx \rho_w$ (at $\alpha_w \approx 1$), and $\rho_s = \alpha_w \rho_w + \alpha_a \rho_a \approx \alpha_w + \rho_a$ (i.e $\rho_s \sim \rho_a$ at $\alpha_w \sim \rho_a$). Therefore, both spray and bubbly liquid layers may be described by more detailed speed and density profiles within a two-fluid air-water system. This, however, increases the system uncertainty, since the spray layer thickness and velocity are unknown. Hence, the spray and bubbly liquids have been excluded from our consideration, while the foam layer yields a basically three-layer system, which can efficiently stabilize the water surface as compared with the two-fluid air-water system. It is of interest that foam bubble sizes $\sim 0.2-2mm$ (Soloviev and Lukas, 2006), which determine the foam interface with air flow, well agree with the experimental correlation for roughness length $\sim 1-2mm$ (Powell et al., 2003).

Finally, note that the zero compressibility and viscosity approximation is commonly accepted in the studies of water surface instabilities in air-water systems (Drazin, 2002; Alexakis et al., 2002). The foam compressibility may be ignored within the same accuracy as that of the air. Indeed, using the smallness of Mach number $M_a = U_a/C_a$ for air and noting that the foam-to-air sound velocity ratio $C_f/C_a \sim \sqrt{\rho_a/\rho_f} \sim \sqrt{\varepsilon}$ (Shtemler and Shreiber, 2006) is of the same order as $U_f/U_a \sim \sqrt{\varepsilon}$ (see Eqs. (18)), it follows that $M_f = U_f/C_f \sim M_a \ll 1$. Although the foam viscosity data in strong wind conditions is, in general, unavailable, artificial foam viscosities are known to be significantly larger than the viscosity of its liquid and gas constituents. On the other hand, natural sea foams are expected to have lower viscosity than their artificial counterpart due to the lack of man-made surfactants and a larger effective size of the foam bubbles (of $\sim 0.2-2mm$). With this in mind, it is assumed that in the range of intermediately short waves under consideration, the viscosity effect on the stability behavior related to the growth rate may be ignored. Ignoring capillary effects (adopted in the present consideration) is valid at the water-foam interface, since the foam is composed from bubbles surrounded by shells of the same sea-water. At the air-foam interface, the surface tension may be naturally assumed equal to that between air and sea-water $\sigma$, and its influence on the water surface stability is rather small, at least, for intermediately short waves with the wavelength $\lambda$: $L_c \ll \lambda \ll L_g$ ($L_c = \sqrt{\sigma/g\rho_w}$), and $L_g = U_a^2/g$ are capillary and gravity lengths (Shtemler et al., 2008). It should be additionally note that accounting for non-Newtonian effects of the foam, which occur due to foam bubbles oscillations (see e.g. Shtemler and Shreiber, 2006) and ignored in the present study, will additionally stabilize the water surface compared with the two-fluid air-water system.

The atmosphere-ocean interaction in strong wind conditions leads to the creation of a foam layer between the atmosphere and the ocean, which provides an effective mechanism of the water surface stabilization, and simultaneously a self–sustained mechanism of foam layer formation due to the destabilization of the air-foam interface. It is conjectured that such stabilization qualitatively explains the observed reduction of roughness and drag.

*Acknowledgments*
Helpful discussions with V. Chernyavskii are gratefully acknowledged.

Modeling, 16, 188-205 .

**Appendix**

The equations of motion that govern the dynamics of the system are the corresponding Euler equations for incompressible fluids in each of the three layers. Besides, kinematic (no voids) and dynamic (pressure continuity) conditions are applied to the foam layer interfaces with water and air:

$$\nabla \cdot \mathbf{v}_j = 0, \quad \rho_j \frac{D\mathbf{v}_j}{Dt} = -\nabla p_j + \rho_j \mathbf{g}, \quad \mathbf{g} = \{0,-g\}$$

$$\mathbf{v}_a = \{U_a, 0\} \quad \text{at} \quad y = \infty, \quad \mathbf{v}_w = \{U_w, 0\} \quad \text{at} \quad y = -\infty.$$

$$\frac{\partial \eta_w}{\partial t} + u_w \frac{\partial \eta_w}{\partial x} - v_w = 0, \quad \frac{\partial \eta_w}{\partial t} + u_f \frac{\partial \eta_w}{\partial x} - v_f = 0, \quad p_w = p_f \quad \text{at} \quad y = \eta_w(x,t),$$

$$\frac{\partial \eta_a}{\partial t} + u_a \frac{\partial \eta_a}{\partial x} - v_a = 0, \quad \frac{\partial \eta_a}{\partial t} + u_f \frac{\partial \eta_a}{\partial x} - v_f = 0, \quad p_a = p_f \quad \text{at} \quad y = \eta_a(x,t). \tag{A1}$$

Here $D/Dt = \partial/\partial t + (\mathbf{v}_j \cdot \nabla)$; $\mathbf{v}_j = \{u_j, v_j\}$; $j = a, f, w$; $\{x, y\}$ are Cartesian coordinates; $t$ is time.

The equilibrium state is now perturbed as follows:

$$f_j(x, y, t) = F_j(y) + F'_j(x, y, t), \quad (j = a, f, w). \tag{A2}$$

Here $f_j$ stands for any of physical variables, $F_j$ and $F'_j$ denote the equilibrium and perturbed values of the velocity and pressure. Then the linearized equations are

$$\frac{\partial U'_j}{\partial x} + \frac{\partial V'_j}{\partial y} = 0, \quad \rho_j \left( \frac{D_j U'_j}{Dt} + \frac{\partial U_j}{\partial y} V'_j \right) = -\frac{\partial P'_j}{\partial x}, \quad \rho_j \frac{D_j V'_j}{Dt} = -\frac{\partial P'_j}{\partial y}, \tag{A3}$$

where $D_j/Dt \equiv \partial/\partial t + U_j \partial/\partial x$, $U_j \equiv const$, $j = a, w, f$. The interface boundary conditions are as follows:

$$\mathbf{V}'_a = \{0,0\} \quad \text{at} \quad y = \infty, \quad \mathbf{V}'_w = \{0,0\} \quad \text{at} \quad y = -\infty,$$

$$\frac{D_w \eta'_w}{Dt} - V'_w = 0, \quad \frac{D_f \eta'_w}{Dt} - V'_f = 0, \quad P'_w - P'_f - g(\rho_w - \rho_f)\eta'_w = 0 \quad \text{at} \quad y = 0,$$

$$\frac{D_a \eta'_a}{Dt} - V'_a = 0, \quad \frac{D_f \eta'_a}{Dt} - V'_f = 0, \quad P'_a - P'_f - g(\rho_a - \rho_f)\eta'_a = 0 \quad \text{at} \quad y = L_f. \tag{A4}$$

Assuming exponential dependence of the wave in $x$ and $t$, one has

$$F'_j(x, y, t) = \tilde{F}'_j(y) \exp(-i\omega t + ikx), \tag{A5}$$

where $k$ and $\omega = kC$ are real and complex wave number and frequency, $C$ is the phase velocity; tilde denotes complex magnitudes. Amplitudes $\tilde{F}_j'$ that satisfy boundary conditions at $y = \pm\infty$ are given by:

$$F_a' = \tilde{F}_a \exp(-ky), \quad F_w' = \tilde{F}_w \exp(ky), \quad F_f' = \tilde{F}_{-f}\exp(-ky) + \tilde{F}_{+f}\exp(ky), \tag{A6}$$

where tilde denotes constant magnitudes. Substitution of relations (A5)-(A6) into the system (A4) yields the piecewise constant equilibrium solution of (A1):

$$ik\tilde{U}_j + \tilde{V}_{jy} = 0, \qquad \rho_j ik\Delta U_j \tilde{U}_j = -ik\tilde{P}_j, \qquad \rho_j ik\Delta U_j \tilde{V}_j = -\tilde{P}_{jy},$$

or, equivalently,

$$i\tilde{P}_j = \frac{\rho_j \Delta U_j}{k}\tilde{V}_{jy}, \qquad \tilde{U}_j = -\frac{1}{\rho_j \Delta U_j}\tilde{P}_j, \qquad \tilde{V}_{jyy} - k^2 \tilde{V}_j = 0, \tag{A7}$$

Here $\Delta U_j = U_j - C$, $j = a,w,f$, $C = \omega/k$ is the phase velocity.

The boundary conditions at infinity and at the foam layer interfaces are

$$\tilde{V}_a = (0,0) \quad \text{at } y = \infty, \qquad\qquad \tilde{V}_w' = (0,0) \quad\quad \text{at } y = -\infty,$$

$$ik\Delta U_w \tilde{\eta}_w - \tilde{V}_w = 0, \quad ik\Delta U_f \tilde{\eta}_w - \tilde{V}_f = 0, \quad \tilde{P}_w - \tilde{P}_f - g(\rho_w - \rho_f)\tilde{\eta}_w = 0 \quad \text{at } y = 0,$$

$$ik\Delta U_a \tilde{\eta}_a - \tilde{V}_a = 0, \quad ik\Delta U_f \tilde{\eta}_a - \tilde{V}_f = 0, \quad \tilde{P}_a - \tilde{P}_f - g(\rho_a - \rho_f)\tilde{\eta}_a = 0 \quad \text{at } y = L_f, \tag{A8}$$

where according to (A7) the eigenfunctions of $\tilde{P}_j$ are related to those of $\tilde{V}_j$ ($j = a,w,f$) as follows:

$$\frac{i\tilde{P}_a}{\tilde{V}_a} = -\rho_a \Delta U_a, \quad \frac{\tilde{U}_a}{\tilde{V}_a} = -i, \quad \frac{i\tilde{P}_w}{\tilde{V}_w} = \rho_w \Delta U_w, \quad \frac{\tilde{U}_w}{\tilde{V}_w} = i, \quad \frac{i\tilde{P}_{\pm f}}{\tilde{V}_{\pm f}} = \pm\rho_f \Delta U_f, \quad \frac{\tilde{U}_{\pm f}}{\tilde{V}_{\pm f}} = \pm i. \tag{A9}$$

Substituting (A9) into the interface boundary conditions (A8), we obtain

$$ik\Delta U_w \tilde{\eta}_w - \tilde{V}_w = 0, \quad ik\Delta U_f \tilde{\eta}_w - \tilde{V}_{-f} - \tilde{V}_{+f} = 0,$$

$$-\rho_a \Delta U_a \tilde{V}_a / \sqrt{E} - \rho_f \Delta U_f \tilde{V}_{+f}\sqrt{E} + \rho_f \Delta U_f \tilde{V}_{-f}/\sqrt{E} - ig(\rho_a - \rho_f)\tilde{\eta}_a = 0,$$

$$ik\Delta U_a \tilde{\eta}_a - \tilde{V}_a/\sqrt{E} = 0, \qquad ik\Delta U_f \tilde{\eta}_a - \tilde{V}_{-f}/\sqrt{E} - \tilde{V}_{+f}\sqrt{E} = 0, \quad E = \exp(2kL_f),$$

$$\rho_w \Delta U_w \tilde{V}_w - \rho_f \Delta U_f \tilde{V}_{+f} + \rho_f \Delta U_f \tilde{V}_{-f} - ig(\rho_w - \rho_f)\tilde{\eta}_w = 0. \tag{A10}$$

The requirement for the discriminant of the linear algebraic system (A10) for amplitudes ($\tilde{F}_j$; $j = a,f,w$) to be equal to zero yields a quartic dispersion relation for phase velocity $C$ (Craik and Adam, 1979; Craik, 1985):

$$2(H_a + H_w) + (E-1)(H_a + 1)(H_w + 1) = 0, \tag{A11}$$

where

$$H_a(C) = \frac{\rho_a(U_a - C)^2 - (\rho_f - \rho_a)g/k}{\rho_f(U_f - C)^2}, \quad H_w(C) = \frac{\rho_w(U_w - C)^2 - (\rho_w - \rho_f)g/k}{\rho_f(U_f - C)^2}.$$

Since the discriminant (A11) of homogeneous linear algebraic system (A10) equals zero, one of the equations in it, e.g. the last equation (A10), should be dropped out, and magnitudes of the eigenfunctions can be expressed through one of the magnitudes, e.g. $\tilde{\eta}_a$, which can be set equal to unity

$$\frac{\tilde{\eta}_w}{\tilde{\eta}_a} = \frac{\sqrt{E}}{2}(H_a + 1) - \frac{1}{2\sqrt{E}}(H_a - 1), \quad \frac{\tilde{V}_w}{\tilde{\eta}_a} = ik\Delta U_w \frac{\tilde{\eta}_w}{\tilde{\eta}_a}, \quad \frac{\tilde{V}_a}{\tilde{\eta}_a} = ik\Delta U_a \sqrt{E},$$

$$\frac{\tilde{V}_{-f}}{\tilde{\eta}_a} = \frac{\sqrt{E}}{2}(H_a + 1)ik\Delta U_f, \quad \frac{\tilde{V}_{+f}}{\tilde{\eta}_a} = -\frac{1}{2\sqrt{E}}(H_a - 1)ik\Delta U_f, \tag{A12}$$

while the magnitudes of the remaining eigenfunctions can be calculated using Eqs. (A9).